\documentclass[global,twocolumn]{svjour}

\usepackage{graphics}
\usepackage{hyperref}
\usepackage{amsmath}
\usepackage{color}
\usepackage[normalem]{ulem}

\begin{document}
\title{Design and fabrication of diffractive atom chips for laser cooling and trapping}
\author{J.\ P.\ Cotter\inst{1,2}\and J.\ P.\ McGilligan\inst{3}\and P.\ F.\ Griffin\inst{3} \and I.\ M.\ Rabey\inst{1}  \and K.\ Docherty\inst{4}\and E.\ Riis\inst{3} \and A.\ S.\ Arnold\inst{3} \and and E.\ A.\ Hinds\inst{1} }
\institute{The Centre for Cold Matter, Blackett Laboratory, Imperial College London, SW7 2AZ, UK 
\and University of Vienna, VCQ, Faculty of Physics, Boltzmanngasse 5, A-1090 Vienna, Austria 
\and Department of Physics, SUPA, University of Strathclyde, Glasgow G4 0NG, UK
\and Kelvin Nanotechnology Ltd, Rankine Building, Oakfield Avenue, Glasgow G12 8LT, UK
}
\date{Received: date / Revised version: date}
\maketitle

\begin{abstract}
It has recently been shown that optical reflection gratings fabricated directly into an atom chip provide a simple and effective way to trap and cool substantial clouds of atoms \cite{nshii13,mcgilligan15}. In this article we describe how the gratings are designed and micro-fabricated and we characterise their optical properties, which determine their effectiveness as a cold atom source. We use simple scalar diffraction theory to understand how the morphology of the gratings determines the power in the diffracted beams. 
\end{abstract}

\section{Introduction}
\label{sec:intro}
Atom chips are microfabricated devices which control and manipulate ultracold atoms in a small, integrated package. Because they provide a convenient way to trap\,\cite{lewis09,pollock09,pollock11}, guide\,\cite{hinds99,dekker00} and detect atoms\,\cite{eriksson05}, atom chips are becoming increasingly important for clocks\,\cite{treutlein04}, Bose-Einstein condensates\,\cite{hansel01,ott01}, matter wave interferometers\,\cite{schumm05,pollock09,baumgartner10}, and quantum metrology\,\cite{riedel10}. In recent years there has been great progress towards integrating a wide range of optical, electric and magnetic elements into atom chips, but the magneto-optical trap (MOT)\,\cite{lindquist92,reichel99} - the element responsible for initial capture and cooling of the atoms - has remained external to the chip.  \begin{figure}[b!]
\centering
\resizebox{\columnwidth}{!}{
\includegraphics{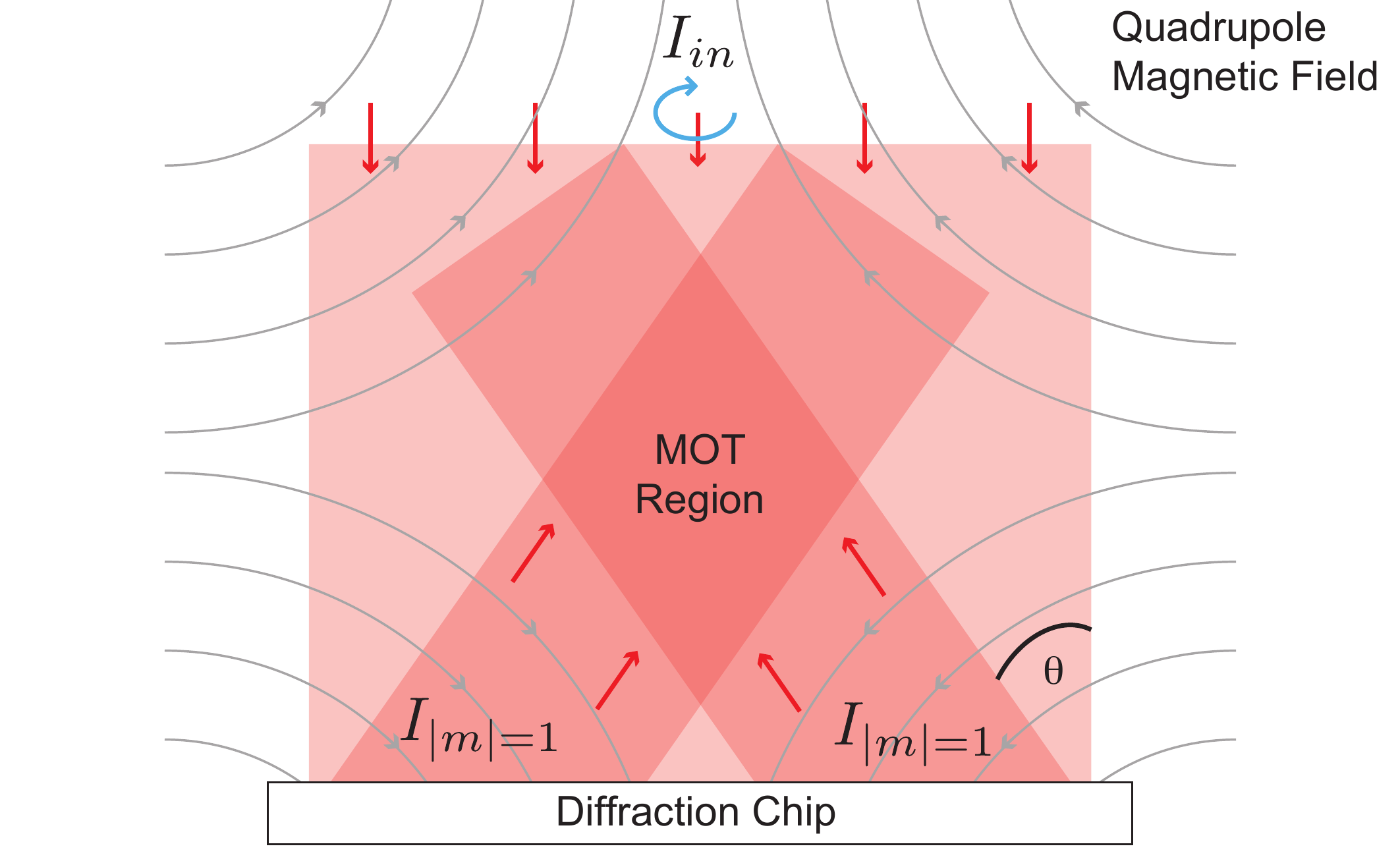}
}
\caption{Principle of the grating chips. A normally-incident laser beam of intensity $I_{in}$ is diffracted by metal reflection gratings, written into the surface of a chip, to make first order beams of intensity $I_{|m| = 1}$. Together, these beams provide the light required for trapping in the magnetic quadrupole field. The angular momentum of the input  beam, indicated by the blue arrow, is opposite to the local magnetic field direction, and the helicity of the light is well preserved after diffraction.
\label{fig:Chip_beams}}
\end{figure}

Following \cite{lee96}, an early attempt to integrate the MOT used deep pyramidal mirrors etched into a thick silicon substrate. These manipulate a single incident laser beam into the overlapping beams required by a MOT. With beams of small size $L$, the number of atoms captured scales as $ L^{6}$\,\cite{pollock11}, a dependence that rolls over to $L^{3.6}$ as the size increases to some centimeters \cite{lindquist92}. The large pyramids favoured by this scaling are not compatible with the normal $500\,\mu$m thickness of a silicon wafer. Although thick wafers are available, days of etching are needed to make pyramids of mm size and additional polishing is required to achieve optical quality surfaces \cite{pollock09,laliotis12,cotter13}. For these reasons the integrated pyramid is unsuitable for applications requiring more than $\sim10^{4}$ atoms. Fig.~\ref{fig:Chip_beams} illustrates a recent extension of this idea where the MOT beams are now formed using microfabricated diffraction gratings, which replace the sloping walls of the pyramid \cite{vangeleyn09,vangeleyn10}. The gratings are easily fabricated on any standard substrate material, and  can readily be made on the centimeter scale. This  allows the MOT to capture up to $10^8$ atoms above the surface of the chip, where they can be conveniently transferred to magnetic traps \cite{hinds99}. Because they only need a small depth of etching, the gratings preserve the  $2$D nature of the structure and sit comfortably with other elements on the chip.  Alternatively, for devices that only require the reliable production of a MOT, the grating chip can be placed outside the wall of a glass cell and used to trap atoms on the inside. 

\begin{figure}[t]
\centering
\resizebox{\columnwidth}{!}{%
\includegraphics{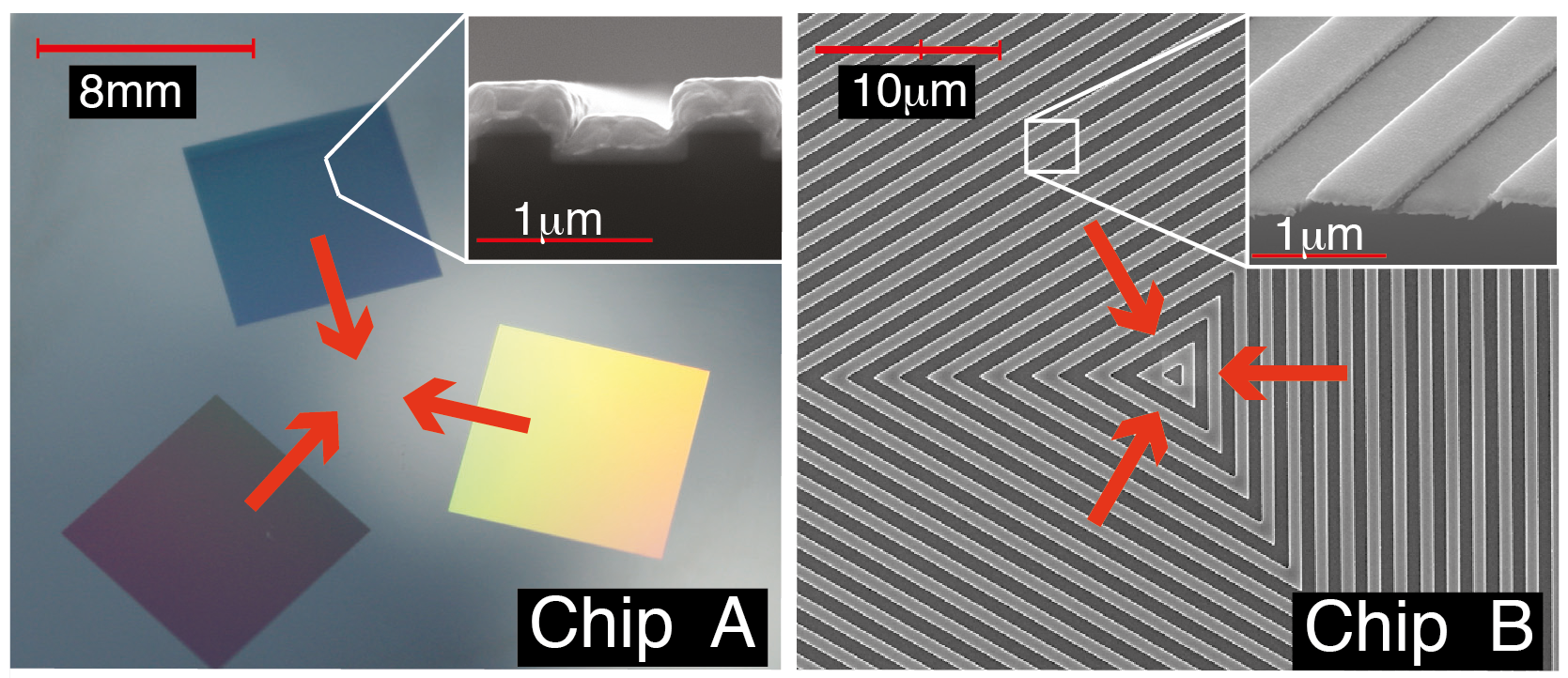}
}
\caption{One-dimensional grating chips of three-fold radial symmetry, used to make 4-beam integrated MOTs. Red arrows indicate the diffracted beams used for trapping. Chip A is made by optical lithography, while chip B (shown magnified) is patterned by e-beam lithography. Insets: Scanning electron microscope images of the grating lines.
\label{fig:SEMs}}
\end{figure}

Figure \ref{fig:SEMs} shows two 1D-grating MOT chips which have already been demonstrated \cite{nshii13}. Chip A has three square grating areas arranged symmetrically to leave a plane area in the centre. Chip B has the same geometry, but the grating pattern covers the whole surface and, in particular, extends all the way to the centre.  In this article we describe the design and fabrication of each chip and compare the expected and measured optical properties of each. The article is organised as follows. In Sec.~\ref{sec:design} we outline the simple scalar diffraction model that we used to design the chips. Section \ref{sec:fab} describes how the gratings were fabricated. In Sec.~\ref{sec:optP} we measure the dimensions of the fabricated gratings and the optical properties of the diffracted beams, and we compare the performance achieved with the theoretical expectations. Finally in Sec.~\ref{sec:conclusion} we summarise our findings.

\section{Design of the chips}
\label{sec:design}

The atoms trapped by the MOT are held by optical scattering forces in the presence of a magnetic quadrupole field. Ideally, these forces should sum to zero at the centre of the quadrupole, which can be achieved by appropriate choices of intensity and polarisation of the light.  The chips described here have symmetry that automatically balances the forces parallel to the surface, but balance in the normal direction has to be designed. Let the incident power $P_{in}$ over an area $A$ of the chip produce power $\eta P_{in}$ in each diffracted beam. The corresponding intensity is $I_{diff}=\eta P_{in}/(A \cos \theta)$, where $\theta$ is the angle to the normal, as shown in Fig.~\ref{fig:Chip_beams}. With $N$ diffracted beams participating in the MOT, the total intensity contributing to the upward force is $N I_{diff} \cos\theta=N \eta P_{in}/A=N\eta I_{in}$. The vertical balance of intensities therefore requires $N \eta =1$. For chips A and B in Fig.~\ref{fig:SEMs}, which use three diffracted beams, this condition becomes $\eta = 1/3$ \cite{vangeleyn10}. In practice, the optimum diffracted intensity is somewhat higher because the polarisations of the upward and downward beams are not the same. 

\begin{figure}[t]
\centering
\resizebox{0.5 \textwidth}{!}{%
 \includegraphics{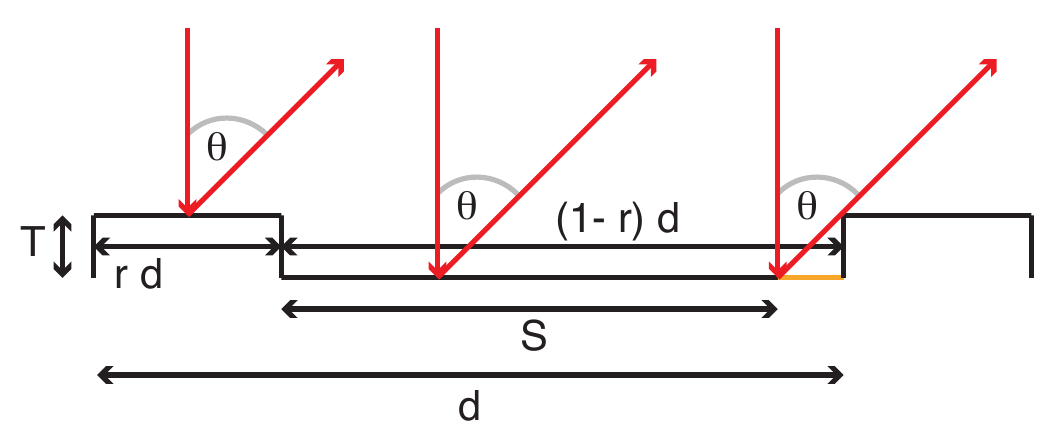}
}
\caption{Ideallised diffraction grating profile, with period $d$, duty factor $r$, and depth $T$. $S$ represents the effective length of the bottom facet, which is shortened because some light is shadowed by the step. Normally incident light is diffracted at an angle $\theta$.   \label{fig:Grat2Source}}
\end{figure}

To estimate the power diffracted from our gratings, we approximate them by the ideal profile shown in Fig.~\ref{fig:Grat2Source}. The elementary period $d$ contains a top face of width $r d$ and a bottom face of width $(1-r) d$ that is lower by a depth $T$. Light diffracted at an angle $\theta$ from the lower face is shadowed by the step, so that the effective width of the face is $S = (1- r) d - T \tan{\theta}$. The phase difference between rays coming from the centre of the top surface and the centre of the effective bottom surface is 
\begin{equation}
\label{eq:phases}
\phi = k \left[ \frac{1}{2} (d - T \tan{\theta})\sin{\theta} - T(1+ \cos{\theta}) \right]\,,
\end{equation}
where $k = 2 \pi /\lambda$ and $\lambda$ is the wavelength of the light. With a normally incident field $E_{in}$, and assuming power reflectivity $\rho$, the diffracted field at (large) distance $R$ is approximated by the Fraunhofer integral.
\begin{figure*}[t!]
\centering
\resizebox{17.5 cm}{!}{%
\includegraphics{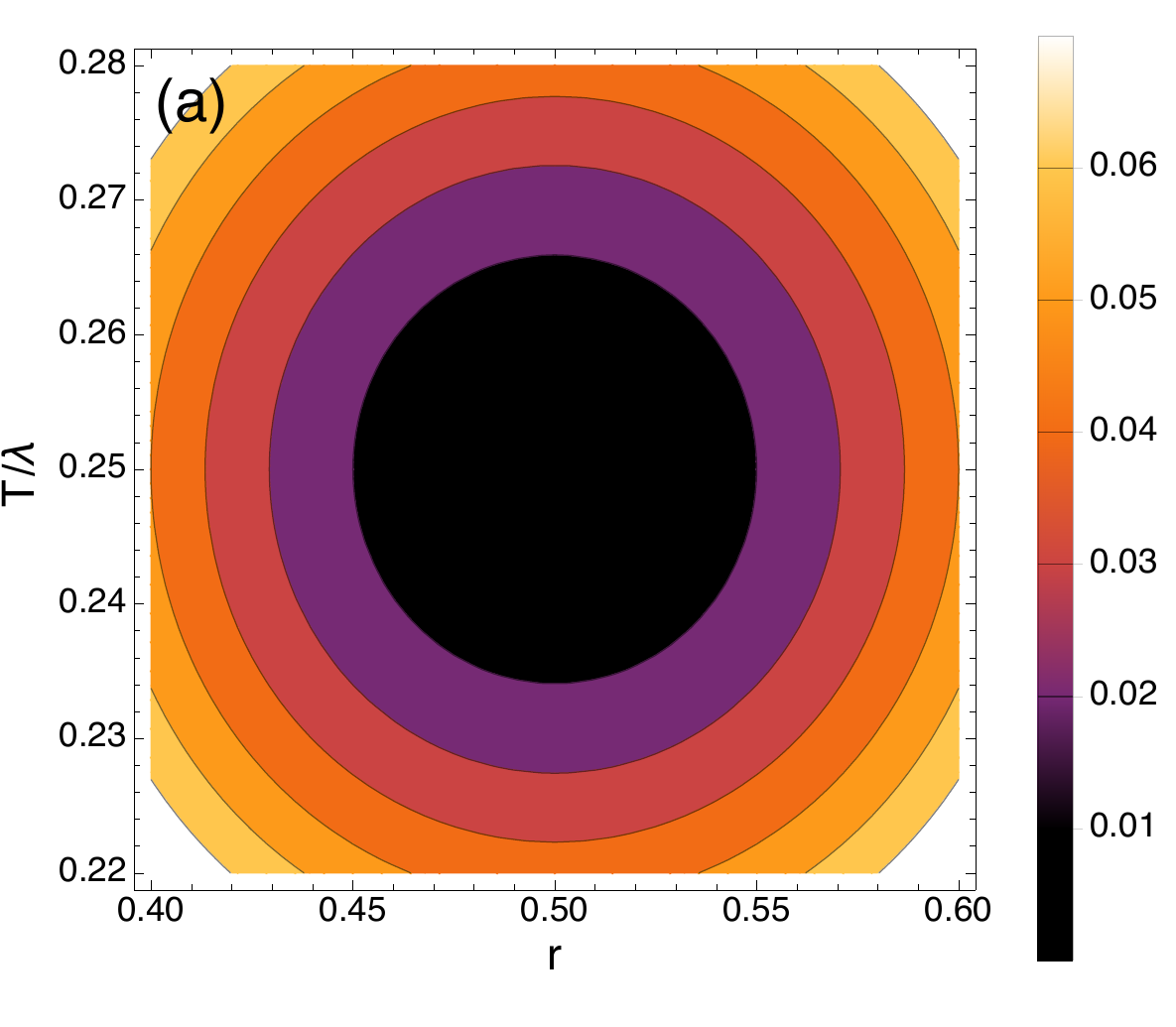}
\includegraphics{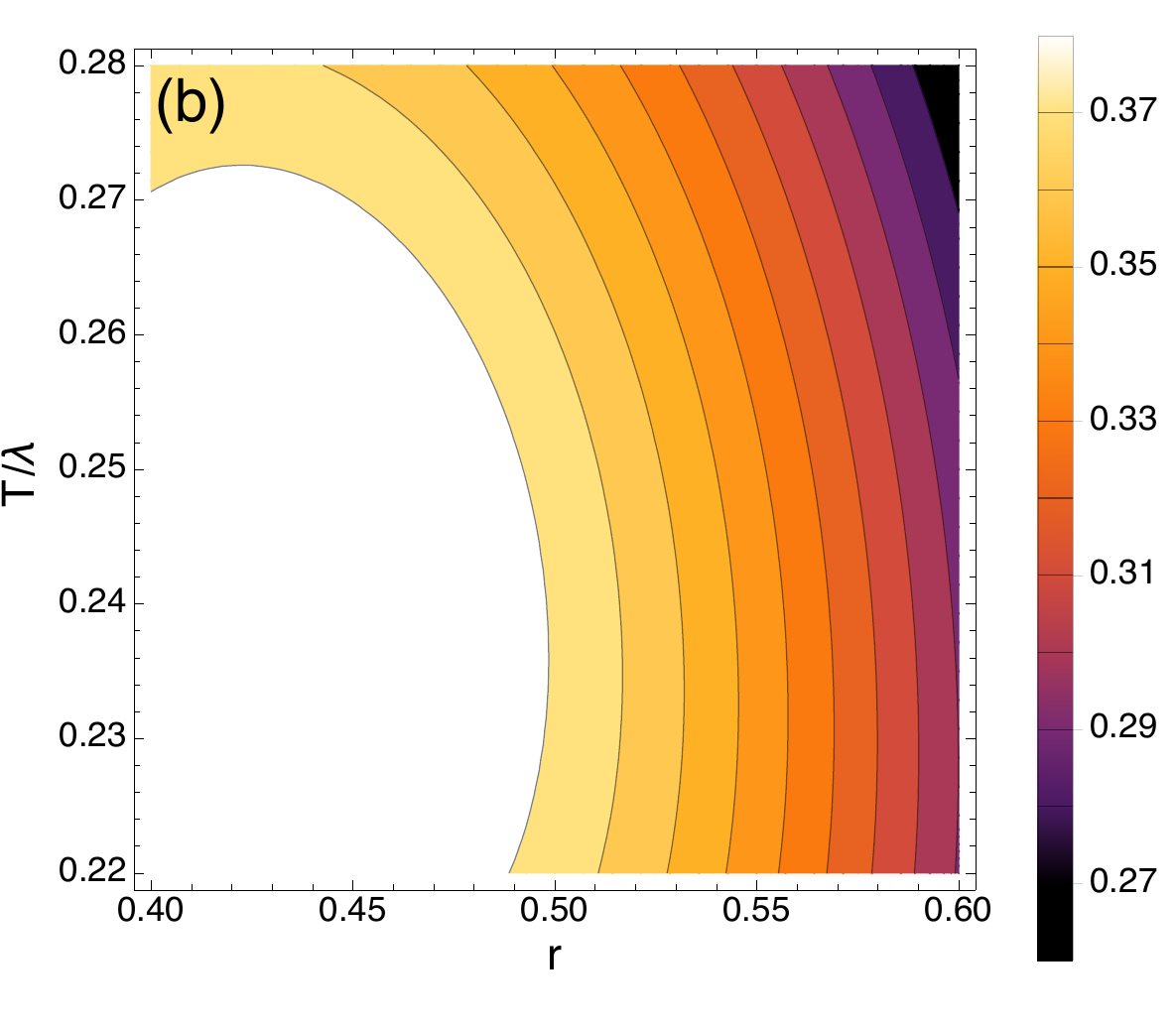}
\includegraphics{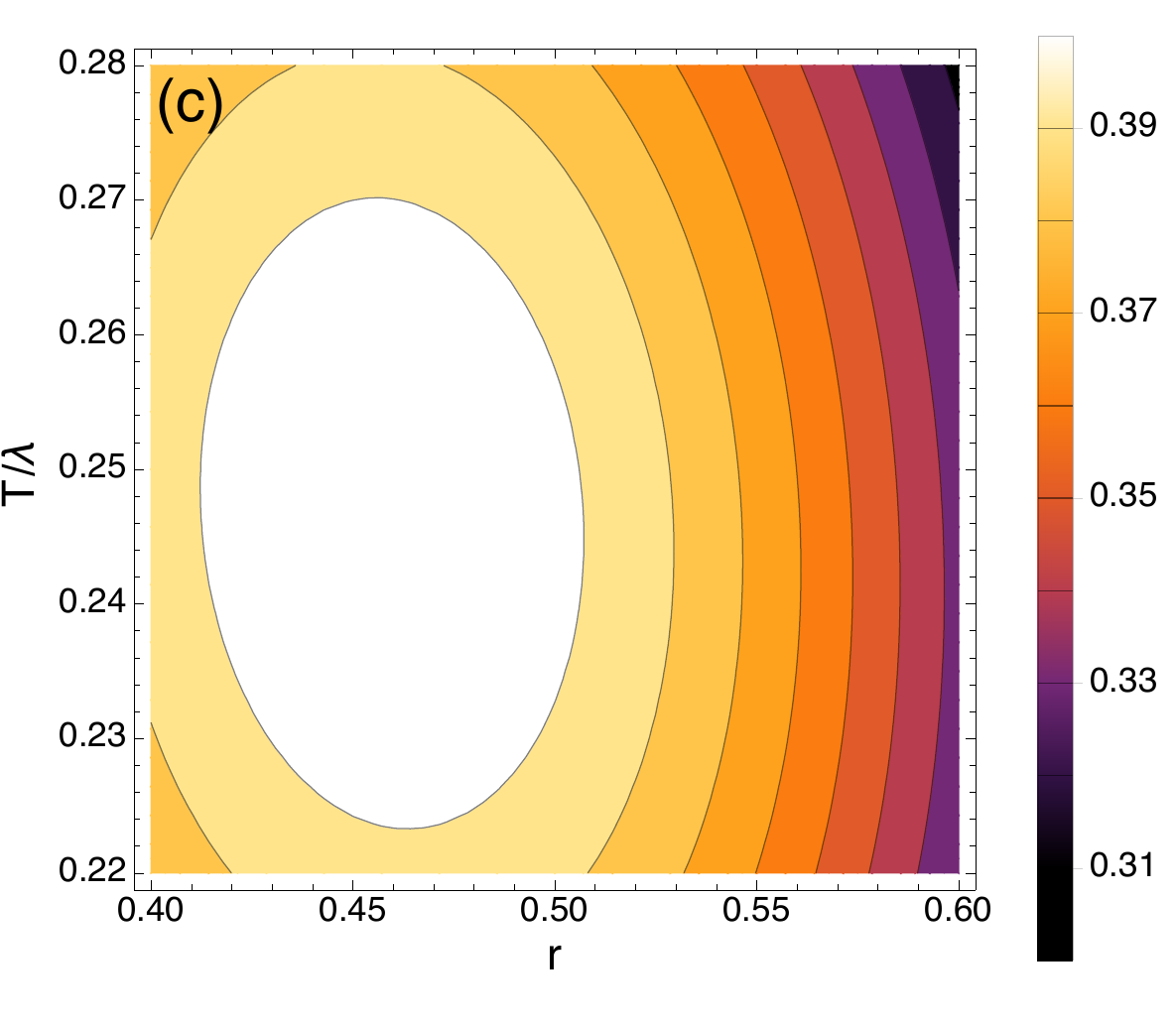}
}
\caption{Power in a single diffraction order, normalised to the incident power and plotted as a function of duty factor $r$ and grating depth $T$ divided by wavelength $\lambda$. Reflectivity is taken to be $\rho=1$. (a) The zero-order case given by Eq.~(\ref{eq:power0}). This is the region near minimum power, where $r\simeq 1/2$ and $T\simeq\lambda/4$. The minimum is wide enough to forgive minor fabrication errors. (b) Fraction of power in the $m=+1$ order of chip A, calculated from Eq.~(\ref{eq:eff}) with $d=1.19\,\mu$m and $\lambda=780\,$nm. (c) Fraction of power in the $m=+1$ order of chip B, calculated from Eq.~(\ref{eq:eff}) with $d=1.48\,\mu$m and $\lambda=780\,$nm. 
\label{fig:m0_2D}}
\end{figure*}

\begin{eqnarray}
\frac{E(\theta)}{E_{in}} &=&\frac{\sqrt{\rho}}{\sqrt{R \lambda}}\left[ \int_{-rd/2}^{rd/2}dx~e^{ikx\sin{\theta}} + e^{i\phi}\int_{-S/2}^{S/2} dx~e^{ikx\sin{\theta}} \right]\nonumber \\
~&~& ~~~~~~~~~~~~~~~~~
\times\left( \sum\limits_{n=1}^{N} e^{iknd\sin{\theta}} \right)\,.
\label{eq:Fraun}
\end{eqnarray}
Here, the first line describes the diffraction from one elementary unit of the grating, as illustrated in Fig.~\ref{fig:Grat2Source}, while the last factor sums over the contribution from all $N$ grating periods. 

The intensity distribution, obtained by squaring equation (\ref{eq:Fraun}), has a comb of narrow peaks coming from the grating factor, with maxima at the Bragg angles given by $\sin{\theta}=m \lambda/d $, where $m$ is an integer. Because many lines of the grating are illuminated, the single-period factor is essentially constant over the small angular spread across one of the Bragg peaks. This makes it straightforward to integrate across the $m^{th}$ Bragg peak to find the total diffracted power $P_{m}$ in that order. The result is

\begin{eqnarray}
\frac{P_m}{P_{in}}  &=& \frac{\rho}{d^2}\left| \int_{-rd/2}^{rd/2}dx~e^{i2\pi m x / d} + e^{i\phi}\int_{-S/2}^{S/2}dx~e^{i2\pi m x / d} \right|^2 \,,\nonumber \\
\end{eqnarray}
$P_{in}$ being the power incident on the $N$ illuminated lines of the grating. Evaluating these integrals,
\begin{eqnarray}
\frac{P_m}{P_{in}} &=& 
\frac{\rho}{m^{2} \pi^{2}} \left[ \right. \sin^{2}{\left( m \pi r \right)} + \sin^{2}{\left( m \pi S/d \right)} \nonumber \\
& ~ & ~~~~~~~~~~ + 2\cos{(\phi)} \sin{\left( m \pi r \right)}\sin{\left( m \pi S/d \right)} \left. \right]. \nonumber \\
&~&
\label{eq:eff}
\end{eqnarray}

Let us first consider diffraction into the $m = 0$ order - i.e. retro-reflection of the incident beam. This needs to be avoided as a strong upward beam of the wrong polarisation is detrimental to the MOT \cite{nshii13}. For chip A there is a plane surface in the central region, which can either be cut away to leave an aperture, or coated with an absorbing layer. For chip B, where the grating structure runs all the way into the middle, the retro-reflection can be suppressed instead by a suitable choice of the grating parameters. On using Eq.~(\ref{eq:phases}) to eliminate $\phi$, Eq.~(\ref{eq:eff}) gives 

\begin{eqnarray}
\label{eq:power0}
\frac{P_{0}}{P_{in}} & = & \rho \left[ 1 + 2 r (r - 1) \left( 1 - \cos{\left(\frac{4 \pi T}{\lambda}\right)} \right) \right].
\end{eqnarray}
This goes to zero when $r = \frac{1}{2}\left( 1 + \frac{i} {\tan{(2 \pi T/\lambda)}} \right)$. Since $r$ must be real we require $\tan{\left(2 \pi T/\lambda\right)}=\infty$, which leaves $r=\tfrac{1}{2}$. It is desirable to minimise the depth $T$ so that $S$ remains as large as possible for the first diffraction order. We therefore choose $T=\lambda/4$. Fig.~\ref{fig:m0_2D}(a) shows how $P_{0}/P_{in}$ varies when $r$ and $T$ deviate from this ideal condition, as they inevitably will in practice. We see that deviations of up to $10\%$ in either $T$ or $r$ give rise to a $P_{0}/P_{in}$ of only one or two percent, making the design robust against minor fabrication errors.

We turn now to the first order beams, which (together with the incident beam) are responsible for making the MOT. The plots in  Fig.~\ref{fig:m0_2D}(b) (for chip A) and Fig.~\ref{fig:m0_2D}(c) (for chip B) show the power $P_1$ in the $m = +1$ order (normalised to $P_{in}$) when the grating depth $T$ and duty factor $r$ are varied. We see that this power is close to a maximum when the retro-reflected power is zero, but can be increased a little by reducing $r$ slightly below 0.5. This has the effect of making $r d$ and $S$ more nearly equal, which improves the contrast of the grating. A little is also gained by reducing $T/\lambda$, so that the width $S$ of the lower surface is increased. As with the minimum of $P_0$, this maximum of $P_1$ is sufficiently forgiving that we are not troubled by minor fabrication errors.

The MOT works because the scattering force in the presence of a magnetic field depends on the polarisation of the light. For that reason, it would be ideal to go beyond this simple scalar model of the diffraction to consider polarisation. However, that theory is quite challenging and  is beyond the scope of this article. Instead we have relied on experiment to determine the polarisation of the diffracted beam, as discussed further in section \ref{sec:optP}.

\section{Fabrication
\label{sec:fab}
}

Chips A and B are produced by two different fabrication methods, which we now describe. 
\subsection{Chip A: Photo-lithography using silicon substrate}

Chip A, shown in Fig.~\ref{fig:SEMs}a, is a $32\,$mm square of silicon in which three $8\,$mm-square lamellar gratings are etched by photolithography. This is then covered with gold to achieve the desired high reflectivity at $780\,$nm. We choose a grating period of $1.2\,\mu$m, which is close to the minimum that can be reliably made by this method. Although we aim for a duty factor of $r=\tfrac{1}{2}$, the bottom face is designed to be $700\,$nm wide, anticipating that $r$ will move towards $\tfrac{1}{2}$ after the gold is added.

\begin{figure}[t]
\centering
\resizebox{\columnwidth}{!}{
\includegraphics{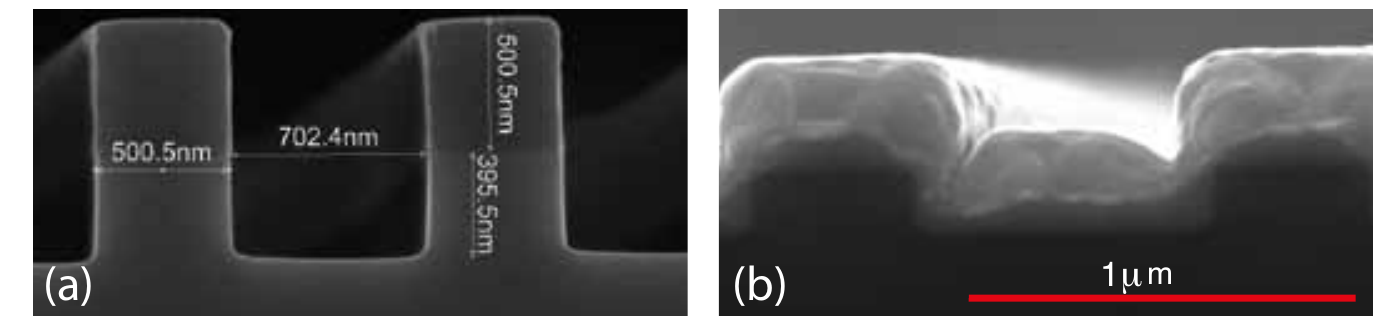}
}
\caption{(a) Scanning electron microscope images of chip A. (a) A deep trench calibrates the etching rate prior to the main fabrication and shows a profile close to that of our model, illustrated in Fig.~\ref{fig:Grat2Source}. (b) The final chip after etching to a depth of  $T \sim 195\,$nm and coating with $200\,$nm of gold. This brings the duty factor $r$ close to $\tfrac{1}{2}$.
\label{fig:IMP_flow}}
\end{figure}

To begin, we make a reticle by direct ebeam writing on chromium-coated quartz.  This is a $5\times$ magnified version of one square grating. A $\langle100\rangle$-orientated $150\,$mm-diameter silicon wafer is then coated with SPR$660$ photoresist to a thickness of 0.8$\,\mu$m and exposed to de-magnified images of the reticle, using light of $365\,$nm wavelength. A stepper motor manoeuvres the reticle to each grating position in turn, to produce an image of $12$ chips - $32$ gratings in total - on the wafer. The resist is then developed, and the exposed silicon is removed by reactive ion etching using an inductively-coupled SF$_{6}$/C$_4$F$_8$ plasma. With a typical etch rate of $\sim5\,$nm/s, this forms a grating of the desired depth - $\lambda/4= 195\,$nm - in under $1\,$minute. The wafer is then stripped of the remaining resist by plasma ashing, before cleaning with a piranha solution to remove any remaining organic contaminants. Figure~\ref{fig:IMP_flow}(a) shows a scanning electron microscope image of a deep grating that was made to calibrate the etch rate. One can see in this image the high quality of the profile and the few-nm accuracy of the widths produced. 

In order to give the gratings a high reflectivity, we apply a $5\,$nm-thick adhesion layer of chromium (by dc sputtering)  followed by $200\,$nm-thick layer of gold (by rf sputtering).  The finished grating is shown in Fig.~\ref{fig:IMP_flow}b. From this and similar scans we measure a final depth of $T = 207(5)\,$nm, a period of $d = 1.19(1)\,\mu$m and a duty factor of $r = 0.51(5)\,$, the latter being due in part to some systematic variation across the chip.

\subsection{Chip B: Electron-beam lithography using silicon substrate}

Chip B is a $22\,$mm square of silicon, coated with aluminium, in which a grating is etched by electron beam lithography.  The grating consists of nested triangles, as shown magnified in Fig.~\ref{fig:SEMs}b, that continue outward to fill a $20\,$mm square. The lamellar surface profile is designed to have a depth of $195\,$nm, a period of $1.5\,\mu$m, and a duty factor of $\tfrac{1}{2}$ . Unlike the photolithography used for chip A, the e-beam fabrication used here is not at all challenged by the resolution we require. However, the large size of the pattern over all does present a challenge.

\begin{figure}[b]
\centering
\resizebox{\columnwidth}{!}{
\includegraphics{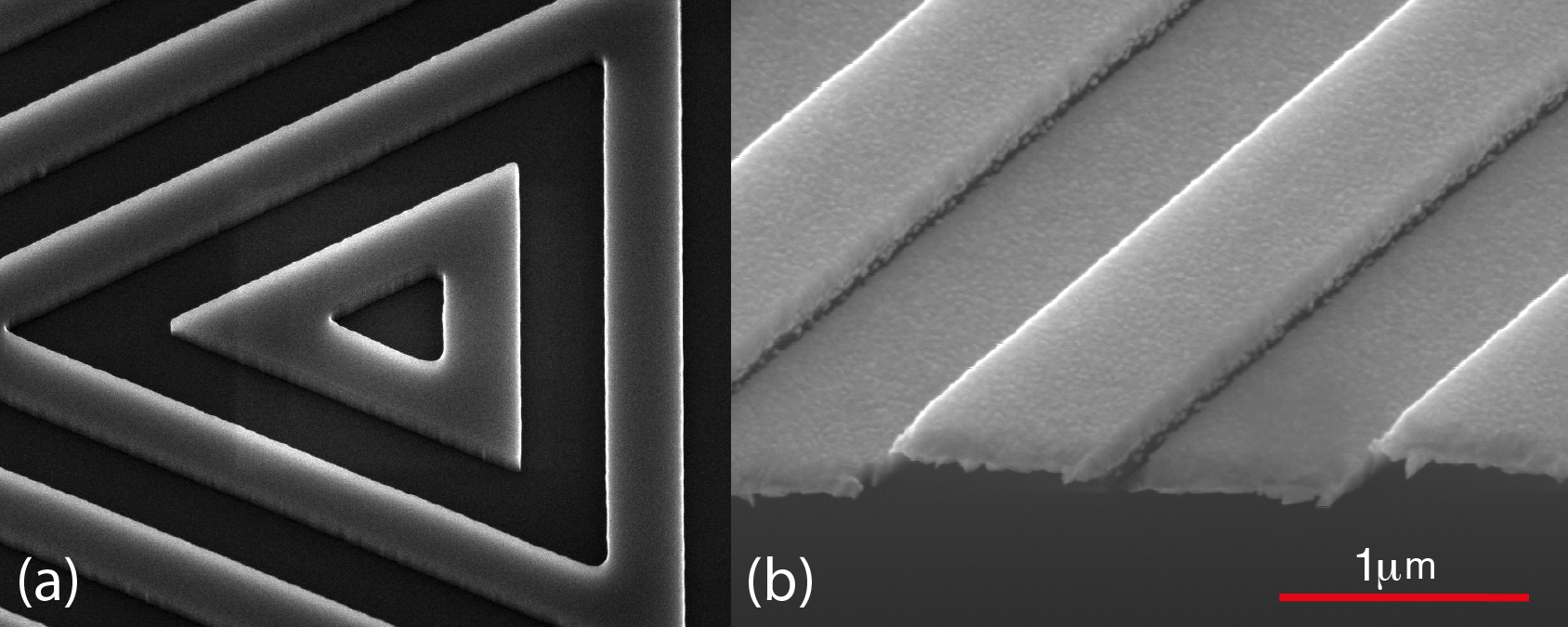}
}
\caption{Scanning electron microscope images of chip B. (a) The centre of chip B, etched to a depth of $195\,$nm, before coating. The triangles are equilateral, but distorted by the angle of view.  (b) After coating with aluminium.
\label{fig:KNT_flow}}
\end{figure}

A $\langle100\rangle$-orientated $100\,$mm-diameter silicon wafer is coated with ZEP520A e-beam resist to a thickness of $350\,$nm, which is then patterned using a high speed e-beam writer (Vistec VB6 with 50MHz scan speed). With $11$ chips, covering a total area of $44\,\mbox{cm}^2$, this takes $25\,$hrs of continuous writing. Particular care is needed to ensure the electron beam direction does not drift over this time, thereby introducing phase variations across individual gratings. The wafer is then etched and cleaned in the same way as chip A. The scanning electron microscope image in Fig.~\ref{fig:KNT_flow}(a) shows the centre of the etched grating and illustrates the high quality of the fabrication. 

After evaporating $100\,$nm of aluminium, the grating is imaged again, as shown in Fig.~\ref{fig:KNT_flow}(b). From this and similar scans we measure the final parameters T = $190(5)$\,nm, d = $1.48(1)\mu$m and r =$ 0.46(5)$.

\section{Measurement of optical properties}
\label{sec:optP}

The reflectivity of each chip was determined by measuring the power in a $780\,$nm laser beam reflected from a flat, un-etched area, and comparing this with the incident power. We found $\rho=0.972(6)$ for chip A and $\rho=0.822(6)$ for chip B. 

In order to measure the diffracted power ratio $P_{m}/P_{in}$, a few-milliwatt laser beam of $780\,$nm wavelength was spatially filtered using a single-mode fibre, then collimated to form a beam of approximately $1\,$mm full-width-half-maximum. This was sent through a polarising beam splitter, then circularly polarised by a quarter-wave plate, as it would be to make a MOT. Roughly $1$\,m from the wave plate, the light was retro-reflected from a flat area of the chip and sent back through the wave plate and beam splitter. The circular polarisation of the incident light was optimised by adjusting the angle of the quarter-wave plate to extinguish the light returning through the beam splitter. Next, a translation stage moved the chip so that the light was incident on a grating, and a power meter then recorded the incident power $P_{in}$ and the power $P_{1}$ diffracted into first order.

We measured each of the three gratings on chip A, with the results $P_1/P_{in}=0.326(2),\,0.323(2)\, 0.386(2)$. These are to be compared with the power ratio given by Eq.~(\ref{eq:eff}) after inserting the measured grating dimensions and reflectivity. That gives $0.340^{+(21)}_{-(36)}$, in good agreement with the measurements. The small variation in both theory and experiment is due predominantly to the variation of $r$. This translates into a variation of the diffracted power because chip A, having $r=0.51(5)$, operates on the high-r side of the maximum plotted in Fig.~(\ref{fig:m0_2D})(b), where the derivative with respect to $r$ is not zero. 

Measurements on the three gratings of chip B gave $P_1/P_{in}=0.381(2),\,0.381(2)\, 0.380(2)$, showing a good level of reproducibility. This is due in part to better uniformity of the e-beam lithography, but also, chip B operates with $r=0.46(5)$, which is very close to the maximum of the plot in Fig.~(\ref{fig:m0_2D})(c), where $P_1$ is insensitive to variation of $r$. The power ratio given by  Eq.~(\ref{eq:eff}) for chip B is $0.328^{+(2)}_{-(9)}$. While this is qualitatively similar to the measured fraction, it does not agree within the measurement uncertainty and we cannot find any plausible adjustment of parameters that might bring them into agreement. We are forced to conclude that our diffraction theory is not able to predict the diffracted power with this high level of accuracy, and suspect that the limitation is due to our use of the effective width $S$, defined by ray optics and therefore not strictly justified.  In the case of chip B, the zeroth order beam passes through the MOT, so it is important with this chip to have a low $P_{0}$. In order to measure this, we rotated the chip by approximately $5\,$mrad to separate the $m=0$ diffracted beam from the incident beam. This measurement gave $P_{0}=0.005(1)$, in good agreement with $0.007^{+(20)}_{-(7)}$ from Eq.~(\ref{eq:eff}).

Because magneto-optical trapping is compromised by the wrong sense of circular polarisation, we looked for this in the first-order diffracted beams using a second combination of quarter-wave plate and polarising beam splitter, adjusted to project the state of the beam onto the basis of left- and right-handed polarisations. Photodetectors at the two beam splitter outputs measured the powers $P_{L}$ and $P_{R}$ in each circular polarisation. The fraction of power in the desired polarisation from the three gratings on Chip A was $88\%$, $ 90\%$ and $98\%$, and we note that better polarisation coincided in each case with higher power. On chip B we measured $97\%$, $ 98\%$ and $99\%$. This high degree of polarisation is more than adequate to make a strong MOT  with either chip \cite{nshii13}. Indeed, although we do not have any calculation for comparison, it seems surprisingly high given the obvious anisotropy of the surface and of the diffraction geometry.  We note that the variation in polarisation is greater across chip A than chip B, and again, we ascribe this to the two different methods of fabrication.

\section{Summary and conclusions}
\label{sec:conclusion}

Optical reflection gratings fabricated on an atom chip offer a simple way to build a large, robust, integrated magneto-optical trap (MOT) for atoms \cite{nshii13}. In this paper we have discussed the main design considerations, and have described how suitable chips can be fabricated using two methods: optical lithography and e-beam lithography. Using scalar Fraunhofer diffraction theory and an idealised model of the lamellar profile, we have provided an account of the expected MOT beam intensities. This theory agrees well with experiment down to the level of a few percent of the incident power, but not with the higher-precision measurements made on the aluminium-coated chip B. We have shown that it is possible to suppress the back-reflection, while at the same time diffracting a large fraction of the power into the two first-order beams. The power in these beams depends on the choice of period $d$, duty factor $r$ and depth $T$ of the grating. These parameters vary a little over the optically fabricated chip A, and rather less over the e-beam fabricated chip B. In either case, we show how to minimise the effect of inhomogeneity on the diffracted beam intensity by operating at the intensity maximum with respect to $r$ and $T$. We also find that the circular polarisation of the light is surprisingly well preserved after diffraction into the first-order beams.

The design principles and theoretical model developed here make this new method accessible to anyone who may wish to incorporate such an integrated trap into an atom chip. We anticipate that this approach will facilitate future quantum technologies using cold and ultra-cold atoms\,\cite{rushton14}.

\section*{Acknowledgments}
The authors acknowledge valuable conversations with Alastair Sinclair of the National Physical Laboratory. This work was supported by the UK EPSRC, ESA (through ESTEC project TEC- MME/2009/66), the CEC FP7 (through project 247687; AQUTE). JPC was funded by an EPSRC support fund and VCQ fellowship, P.G. by the Royal Society of Edinburgh and E.H. by the Royal Society.

\bibliographystyle{unsrt}

\begin{thebibliography}{99}

\bibitem{nshii13}
C.~C. Nshii, M.~Vangeleyn, J.~P. Cotter, P.~F. Griffin, E.~A. Hinds, C.~N.
  Ironside, P.~See, A.~G. Sinclair, E.~Riis, and A.~S. Arnold.
\newblock A surface-patterned chip as a strong source of ultracold atoms for
  quantum technologies.
\newblock {\em nature nanotechnology}, 8:321 Ð-- 324, 2013.

\bibitem{mcgilligan15}
J.~P. McGilligan, P.~F. Griffin, E.~Riis, and A.~S. Arnold.
\newblock Phase-space properties of magneto-optical traps utilising
  micro-fabricated gratings.
\newblock {\em Opt. Express}, 23(7):8948--8959, Apr 2015.

\bibitem{lewis09}
G.~N. Lewis, Z.~Moktadir, C.~Gollasch, M.~Kraft, S.~Pollock,
  F.~Ramirez-Martinez, J.~Ashmore, A.~Laliotis, M.~Trupke, and E.~A. Hinds.
\newblock Fabrication of magnetooptical atom traps on a chip.
\newblock {\em J. MEMS}, 18:347 -- 353, 2009.

\bibitem{pollock09}
S.~Pollock, J.~P. Cotter, A.~Laliotis, and E.~A. Hinds.
\newblock Integrated magneto-optical traps on a chip using silicon pyramid
  structures.
\newblock {\em Optics Express}, 17:14109 -- 14114, 2009.

\bibitem{pollock11}
S.~Pollock, J.~P. Cotter, A.~Laliotis, F.~Ramirez-Martinez, and E.~A. Hinds.
\newblock Characteristics of integrated magneto-optcal traps for atom chips.
\newblock {\em New J. Phys.}, 13:043029, 2011.

\bibitem{hinds99}
E.~A. Hinds and I.~G. Hughes.
\newblock Magnetic atom optics: mirrors, guides, traps, and chips for atoms.
\newblock {\em Journal of Physics D: Applied Physics}, 32(18):R119, 1999.

\bibitem{dekker00}
N.~H. Dekker, C.~S. Lee, V.~Lorent, J.~H. Thywissen, S.~P. Smith,
  M.~Drndi\'{c}, R.~M. Westervelt, and M.~Prentiss.
\newblock Guiding neutral atoms on a chip.
\newblock {\em Phys. Rev. Letts.}, 84:1124, 2000.

\bibitem{eriksson05}
S.~Eriksson, M.~Trupke, H.~F. Powell, D.~Sahagun, C.~D.~J. Sinclair, E.~A.
  Curtis, B.~E. Sauer, E.~A. Hinds, Z.~Moktadir, C.~O. Gollasch, and M.~Kraft.
\newblock Integrated optical components on atom chips.
\newblock {\em Eur. Phys. J. D}, 35:135 -- 139, 2005.

\bibitem{treutlein04}
P.~Treutlein, P.~Hommelhoff, T.~Steinmetz, T.~W. H\"{a}nsch, and J.~Reichel.
\newblock Coherence in microchip traps.
\newblock {\em Phys. Rev. Letts.}, 92:203005, 2004.

\bibitem{hansel01}
W.~H\"{a}nsel, P.~Hommelhoff, T.~W. H\"{a}nsch, and J.~Reichel.
\newblock Bose-{E}instein condensation on a microelectronic chips.
\newblock {\em Nature}, 413:498--501, 2001.

\bibitem{ott01}
H.~Ott, J.~Fort\'{a}gh, F.~Schlotterback, A.~Grossmann, and C.~Zimmermann.
\newblock Bose-{E}instein condensation in a surface microtraps.
\newblock {\em Phys. Rev. Letts.}, 23:230401, 2001.

\bibitem{schumm05}
T.~Schumm, S.~Hofferberth, L.~M. Andersson, S.~Wildermuth, S.~Groth,
  I.~Bar-Joseph, J.~Schmiedmayer, and P.~Kr\"{u}ger.
\newblock Matter-wave interferometry in a double well on an atom chip.
\newblock {\em Nat. Phys.}, 1:57--62, 2005.

\bibitem{baumgartner10}
F.~Baumg\"{a}rtner, R.~J. Sewell, S.~Eriksson, I.~Llorente-Garcia, J.~Dingjan,
  J.~P. Cotter, and E.~A. Hinds.
\newblock Measuring energy differences by {BEC }interferometry on a chips.
\newblock {\em Phys. Rev. Letts.}, 105:243003, 2010.

\bibitem{riedel10}
M.~F. Riedel, P.~B\"{o}hi, Y.~Li, T.~W H\"{a}nsch, A.~Sinatra, and
  P.~Treutlein.
\newblock Atom-chip-based generation of entanglement for quantum metrology.
\newblock {\em Nature}, 464:1170, 2010.

\bibitem{lindquist92}
K.~Lindquist, M.~Stephens, and C.~Wieman.
\newblock Experimental and theoretical-study of the vapor-cell zeeman optical
  trap.
\newblock {\em Phys. Rev. A}, 46(7):4082--4090, 1992.

\bibitem{reichel99}
J.~Reichel, W.~H\"ansel, and T.~W. H\"ansch.
\newblock Atomic micromanipulation with magnetic surface traps.
\newblock {\em Phys. Rev. Lett.}, 83:3398--3401, Oct 1999.

\bibitem{lee96}
K.~I. Lee, J.~A. Kim, H.~R. Noh, and W.~Jhe.
\newblock Single-beam atom trap in a pyramidal and conical hollow mirror.
\newblock {\em Opt. Letts.}, 21:1177, 1996.

\bibitem{laliotis12}
A.~Laliotis, M.~Trupke, J.~P. Cotter, G.~Lewis, M.~Kraft, and E.~A. Hinds.
\newblock {ICP} polishing of silicon for high-quality optical resonators on a
  chip.
\newblock {\em Journal of Micromechanics and Microengineering}, 22(12):125011,
  2012.

\bibitem{cotter13}
J~P Cotter, I~Zeimpekis, M~Kraft, and E~A Hinds.
\newblock Improved surface quality of anisotropically etched silicon $\{111\}$
  planes for mm-scale optics.
\newblock {\em Journal of Micromechanics and Microengineering}, 23(11):117006,
  2013.

\bibitem{vangeleyn09}
M.~Vangeleyn, P.F. Griffin, E.Riis, and A.~S. Arnold.
\newblock Single-laser, one beam, tetrahedral magneto-optical trap.
\newblock {\em Opt. Express}, 17(16):13601--13608, 2009.

\bibitem{vangeleyn10}
M.~Vangeleyn, P.~F. Griffin, E.~Riis, and A.~S. Arnold.
\newblock Laser cooling with a single laser beam and a planar diffractor.
\newblock {\em Opt. Lett.}, 35(20):3453--3455, Oct 2010.

\bibitem{rushton14}
J.~A. Rushton, M.~Aldous, and M.~D. Himsworth.
\newblock Contributed review: The feasibility of a fully miniaturized
  magneto-optical trap for portable ultracold quantum technology.
\newblock {\em Review of Scientific Instruments}, 85(12), 2014.

\end{thebibliography}

\end{document}